\documentclass[10pt,pop,aip,twocolumn,superscriptaddress]{revtex4-1}
\usepackage[utf8]{inputenc}
\usepackage{amsmath}
\usepackage{graphicx,color}
\usepackage{mathptmx}
\usepackage{helvet}

\begin{document}

\title{Effects of high energy photon emissions in laser generated ultra-relativistic plasmas: real-time synchrotron simulations}

\author{Erik Wallin} 
\affiliation{Department of Physics, Ume{\aa} University, SE--901 87 Ume{\aa}, Sweden}
\affiliation{Department of Applied Physics, Chalmers University of Technology, SE--412 96 G\"oteborg, Sweden}
\author{Arkady Gonoskov} 
\affiliation{Department of Applied Physics, Chalmers University of Technology, SE--412 96 G\"oteborg, Sweden}
\affiliation{Institute of Applied Physics, Russian Academy of Sciences, Nizhny Novgorod 603950, Russia}
\affiliation{University of Nizhny Novgorod, Nizhny Novgorod 603950, Russia}
\author{Mattias Marklund}
\affiliation{Department of Applied Physics, Chalmers University of Technology, SE--412 96 G\"oteborg, Sweden}

\begin{abstract}
We model the emission of high energy photons due to relativistic charged particle motion in intense laser-plasma interactions. This is done within a particle-in-cell code, for which high frequency radiation normally cannot be resolved due to finite time steps and grid size. A simple expression for the synchrotron radiation spectra is used together with a Monte-Carlo method for the emittance. We extend previous work by allowing for arbitrary fields, considering the particles to be in instantaneous circular motion due to an effective magnetic field. Furthermore we implement noise reduction techniques and present validity estimates of the method. Finally, we perform a rigorous comparison to the mechanism of radiation reaction, and find the emitted energy to be in excellent agreement with the losses calculated using radiation reaction.
\end{abstract}

\maketitle

\section{Introduction}
The evolution of laser systems has prompted an evolution in the modeling of laser-matter interactions, in particular ultra-intense laser-matter interactions. Up until now, laser generated plasmas have behaved in a classical manner (with a few exceptions \cite{birdsall1991particle,ridgers2014modelling,chen2013modeling}), and they have been successfully modelled using so called particle-in-cell (PIC) schemes. \cite{Birdsall1985,Dawson1983} However, with planned upgrades of current laser systems, as well as new large-scale facilities, \cite{vulcan, eli,xcels} there is a need to push the modeling of laser-matter systems even further. It is expected that semi-classical and/or quantum electrodynamical (QED) effects could have a significant effect on the interaction between ultra-intense lasers and matter targets. Merging these semi-classical or QED effects with the classical codes presents a significant challenge, in particular since (a) the classical and the quantum systems have different mathematical setups, (b) there are still scarce experimental data to compare with, and (c) the classical and QED scales are very different. Thus, great care has to be taken when modeling, e.g., the quantum version of radiation reaction. \cite{ilderton2013RRinStrongQED,harvey2015testing} However, before the onset of a fully "quantum" behavior of these systems, there are regimes where we can have significant emission of radiation in terms of high frequency photons. Such emissions can be interesting in their own right, but also constitutes a testbed for efficient computational models. 

The traditional PIC scheme is  well suited for the simulation of lasers interacting with classical plasmas. However, the future high-power laser regime will have the capacity of producing ultra-relativistic particles radiating with frequencies that the PIC grid will be unable to resolve. 

With the particles seen to be in instantaneous circular motion with the frequency $\omega_H$ they will emit high frequency synchrotron radiation with a typical frequency\cite{LL.V2, jackson1998classical}  $\omega_c = \frac{3}{2} \omega_H \gamma^3$.  

Thus, the typical frequency $\omega_c$ grows rapidly as we increase the $\gamma$-factor. This occurs as the radiation reaction force on the particles begin to be of importance, and the energy loss of the particles due to this could be seen as an account of the radiated energy leaking into the unresolvable part of the spectrum.

Here we present a simple runtime algorithm for emitting high frequency radiation in the form of particles. We extend previous work \cite{Rousse2004} by considering the particles to be in instantaneous circular motion due to an effective magnetic field, thus accounting for acceleration from arbitrary fields. This gives a simple and computationally scalable method for real-time emission of highly energetic photons in a PIC simulation.

\section{The field setup}
The traditional PIC approach treats a plasma as an ensemble of charged particles moving in an electromagnetic (EM) field defined on a grid. The radiation from these particles is accounted for via current densities defined on the grid. In such way, using the equations of motion for the particles and Maxwell's equations for the EM-field, we can describe the energy circulation between the particles and the EM-field in a self-consistent way.
 
This approach, however, only takes into account the radiation that is resolved by the grid. Radiation with frequency above $\omega_{grid} = c/\Delta x$, where $\Delta x$ is the distance between grid points and $c$ is the light velocity, cannot be resolved by the grid as the wavelength then is smaller than the sampling distance.

As we increase the laser intensity, the particles can reach higher energies. This will extend the spectral range of the synchrotron emission, whose typical frequency scales as \cite{LL.V2}
\begin{equation}
\omega_c = \frac{3eH_{\text{eff}}}{2mc} \gamma^2,
\end{equation}
where $e$ and $m$ are electron charge and mass, respectively, $\gamma$ is the electron gamma factor, and $H_{\text{eff}}$ is the \emph{effective magnetic field}. This is defined as the magnetic field that can cause the same transverse acceleration as an electron experiences (from the combined electric and magnetic fields). One could think that this inevitably raise the requirements on the grid's resolution and essentially ruins the possibility for simulation in the ultra-relativistic case.

Fortunately, it is possible to consider the EM-energy as occupying two well-separated frequency domains \cite{QEDPIC}. The first region (high frequency) is associated with the individual synchrotron emission of an electron, where as the second (low frequency) is associated with the coherent emission from many electrons. We can then model the high frequency emission with particles (photons), in addition to the classical fields.

\section{Synchrotron emission}
To calculate the classical emission spectra from particles in a simulation, the natural starting point is the expression 
\begin{equation}\label{em_int}
    \frac{\mathrm{d}^2 I}{\mathrm{d}\omega \mathrm{d} \Omega} = \frac{e^2}{4\pi^2 c} \Bigg| \int_{-\infty}^{\infty} \!\! \frac{\boldsymbol{n}\times [(\boldsymbol{n}-\boldsymbol{\beta})\times \dot{\boldsymbol{\beta}}]}{(1-\boldsymbol{\beta}\cdot \boldsymbol{n})^2} e^{i\omega(t-\frac{\boldsymbol{n}\cdot\boldsymbol{r}(t)}{c})} \mathrm{d} t \Bigg|^2 
\end{equation}
for the radiation from an accelerated particle \cite{jackson1998classical}, 
where $\boldsymbol{r}$ is the position of the particle, $\boldsymbol{\beta}=\boldsymbol{v}/c$ the velocity and $\boldsymbol{n}$ is the direction to the observer. Here the integral of the particle path during all times must be calculated, e.g. using data from a PIC-simulation for the particle paths. Assuming smooth paths for the particle in between time steps one can calculate the high frequency part of the spectra despite the Nyqvist-Shannon sampling theorem \cite{thomas2010algorithm,chen2013modeling} (the end-point effect in the integral is studied in \cite{chen2013modeling}). These post-processing methods all have the setback of not being able to execute at runtime, which is a demand if we wish to be able to inject photons as particles in a PIC simulation. Furthermore it is a detailed way of calculating the emission which often exceeds our needs.

An evident alternative is to calculate the integral in Eq. (\ref{em_int}) real-time for each particle. To do this we must neglect some interference between contributions from different time steps \cite{harvey2015testing}, as Eq. (\ref{em_int}) involves the square of the integral over all times. More disconcerting, this would be a very computationally costly method.

To reduce and control the computational costs we will present a statistical routine where we calculate the contribution from each time step and particle separately, using a simplified expression for the radiation emission. This procedure implies sacrificing the interference between emissions during different time steps (e.g. neighbouring ones), which in principle could pose a problem. However, it turns out that in regimes typical for laser-plasma interactions with sufficient laser intensity (i.e., when radiation reaction losses are important), we can neglect the interference between neighbouring time steps.

To understand the potential problem, and the accuracy of this procedure, let us consider the  case of an electron passing through the alternating fields of an undulator. As is well known the emission can have a synchrotron spectrum, spreading up to $\hbar \omega_{max} \sim \gamma^3$, (wiggler regime) for $K \gg 1$ or an undulator spectrum, spreading up to  $\hbar \omega_{max} \sim \gamma^2$, (undulator regime) for $K < 1$, where $K = \theta/\gamma^{-1}$, (here $\theta$ denotes the maximum angle between local and global propagation). The factor $\gamma^2$ in the undulator regime originates from a Lorentz transformation, while in the wiggler regime there is one more $\gamma$-factor due to the radiation being concentrated within the angle $\gamma^{-1}$ along the direction of local propagation, which for an observer appears as a train of pulses with duration proportional to $\gamma^{-1}$.

As seen from these cases, interference of radiation from neighbouring time steps can dramatically change the spectrum. For this problem the procedure under consideration is not able to distinguish if the emission has a synchrotron or an undulator spectrum based solely on the current time step. However, in terms of "arbitrary" EM-fields, as in laser-plasma interaction, we can assume the wiggler regime to be valid, if the time scale $\tau_{EM}$ of the EM-field variations is larger than the time of the particle rotation by an angle $\gamma^{-1}$, which can be written as $\tau_{EM} > \gamma^{-1}\omega_H^{-1}$, where $\omega_H = eH_{\text{eff}}/mc\gamma$ is the effective cyclotron frequency for the local motion. Using the estimates $H=a mc \omega_L/e$ (here $a \sim \gamma$ is the field amplitude in relativistic units) we can rewrite this constraint inequality, and thus assume a synchrotron spectrum when
\begin{equation}
  a^{-1} < \omega_L \tau_{EM}.
\end{equation}
Radiation reaction losses becomes notable for $a > 100$ for micron wavelength lasers, which implies $\omega_L\tau_{EM} > 10^{-2}$. Thus we conclude that the emitted radiation can have a non-synchrotron spectrum only if the radiating process provides a significant energy conversion to the 100th harmonic, which is a rather exotic case. Thus, for many laser applications it is reasonable to neglect interference between the radiation emitted at different time steps and assume synchrotron type of emission. However, this approximation need to be kept in mind. 

For the above case, the recoil of the particles can be described by the last term of the Landau-Lifshitz force \cite{LL.V2}
\begin{equation}
  \label{eq:LLformOfRRForce}
  \textbf{f}_{RR} = -\frac{2}{3}\frac{e^2 m^2 c}{\hbar^2} \chi^2 \textbf{v},
\end{equation}
where $\textbf{v}$ is the particle's velocity and $\chi = \frac{2}{3} \hbar \omega_c / \gamma mc^2$ i the ultra-relativistic case. This last parameter is a measure of the typical emitted photon energy over the particle energy, and for $\chi > 1$ a QED treatment of the problem is needed \cite{ridgers2014modelling}. The radiation reaction force has previously been implemented in PIC codes \cite{Zhidkov2002,Nakamura2012}. Here we focus at the accurate description of the related synchrotron emission.

\section{Modelling the relativistic case}
We are interested in simulating the emission from relativistic particles formed in laser-plasma interaction. Furthermore, our focus is on the unresolvable (as compared to the computational scale parameters) high frequency part of the spectrum, i.e., high energy photons. The emitted power due to the \emph{transverse} acceleration of a particle is a factor $\gamma^2$ larger than the emitted power due to the \emph{longitudinal} acceleration \cite{jackson1998classical}. Thus, for high energy electrons the main part of the emitted radiation is due to synchrotron radiation ($\mathbf{a} \perp \mathbf{v}$). Likewise, the main source of high energy photons will be due to the transverse acceleration of high energy electrons. 

We will consider the particles to be in instantaneous circular motion and, from their position and momentum during two neighbouring time steps, determine the corresponding \emph{effective} magnetic field $H_{\mathrm{eff}}$. From this and the particle energy the corresponding synchrotron spectrum is determined from which we simulate the emission using a Monte-Carlo method.

The energy radiated per unit frequency interval and unit solid angle for a relativistic particle moving in an instantaneous circular motion is given by \cite{jackson1998classical}
\begin{align}
  \frac{\mathrm{d}^2 I}{\mathrm{d}\omega \mathrm{d} \Omega} =& \frac{e^2}{3\pi^2 c} \left( \frac{\omega \rho}{c}\right)^2 \left( \frac{1}{\gamma^2} + \theta^2 \right)^2 \notag \\
&\times \left[ K_{2/3}^2(\xi) + \frac{\theta^2}{(1/\gamma^2) + \theta^2} K_{1/3}^2(\xi) \right] ,
\end{align}
where $\rho$ is the radius of curvature, $\theta$ the angle from the plane of rotation, $K_{1/3}$ and $K_{2/3}$ are modified Bessel functions, and 
\begin{equation}
  \xi = \frac{\omega \rho}{3c} \left( \frac{1}{\gamma^2} + \theta^2\right)^{3/2}.
\end{equation}
For strongly relativistic particles the radiation is confined to small angles ($\sim 1/\gamma$) around the direction of propagation of the particle. This is particularly true for the high frequency radiation that is our prime interest. 

To derive a method with computational economy, we make use of the approximation that the emitted high energy photons are along the direction of motion of the emitting particle \cite{Rousse2004}. We can then integrate the emission spectra over all angles to get an expression for the intensity of synchrotron radiation as a function of frequency only, which can be written as \cite{LL.V2}
\begin{equation} 
  \label{eq:synchrotron}
 \frac{\partial I}{\partial \omega} = \frac{\sqrt{3}}{2\pi} \frac{e^3 H_{\text{eff}}}{mc^2} F\left(\frac{\omega}{\omega_c}\right) ,
\end{equation}
where
\begin{equation}
  F\left(\xi\right) = \xi \int^{\infty}_{\xi} K_{5/3}\left(\xi\right) d\xi \, .
\end{equation}
These expressions gives us the basis for setting up your computational scheme.

\section{Numerical implementation}
The method of synchrotron photon emission is implemented in the PIC codes ELMIS3D and Picador, \cite{QEDPIC, bastrakov.jcs.2012} using their shared ``module-development-kit''. 

The force $\mathrm{d} \mathbf{p}/\mathrm{d} t$ on the particle from the fields is known and we calculate its average momentum $\mathbf{p}$. Considering the part of the change in momentum which is perpendicular to the momentum we can calculate the magnitude of the effective magnetic field $H_{\text{eff}}$. As this field is perpendicular to the plane of rotation, its magnitude is given by 
\begin{equation}
  \label{eq:effective_H}
  H_{\text{eff}} = \frac{m c \gamma}{e} \frac{1}{\Delta t} \frac{\sqrt{\mathbf{p}^2 \mathrm{d}\mathbf{p}^2 - (\mathbf{p} \cdot \mathrm{d}\mathbf{p})^2 }}{p^2}.
\end{equation}
In the expression above the assumption is made that the change in momentum is smaller than the average momentum. This is in general a good assumption, but situations can still occur in simulations in which it is violated. In the Boris scheme \cite{boris1970relativistic} for changing the momentum of the particles, half the electric force is first applied, then a rotation due to the magnetic field is performed, and finally the remaining half of the electric force can be applied. If the momentum of a particle is greatly reduced by the electric field in the first step then the magnetic field could turn it around completely and finally it could be strongly accelerated "backwards" due to the remaining part of the electric field. This would produce a change in the momentum much greater than the average momentum and give a very large value of the effective magnetic field, where in practice a much smaller magnetic field was changing the direction of a much less energetic particle. This would give rise to noise in form of singular high-energy photon emission. This is resolved by imposing the upper limit $\Delta p_{\perp}/p < 1$ in Eq. (\ref{eq:effective_H}). 

\subsection{Monte Carlo method}
\begin{figure}
  \includegraphics[width=0.5\textwidth]{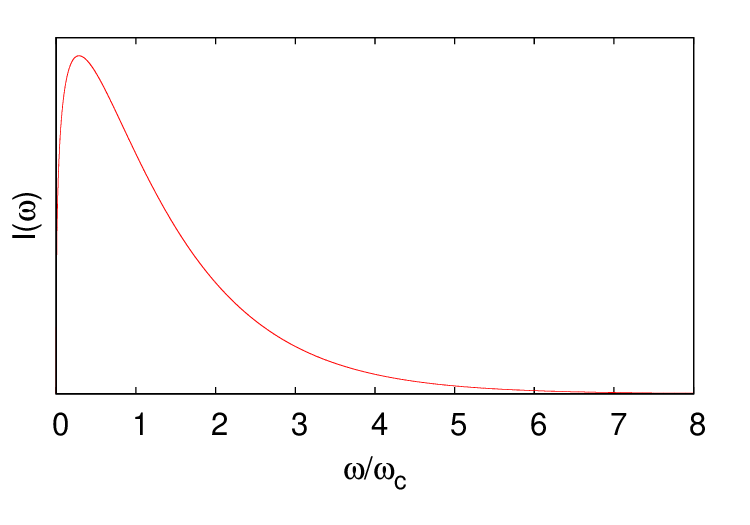}
  \caption{Emitted energy as a function of frequency for an ultra-relativistic particle in circular motion. Typical frequency of radiation, $\omega_c$ is given by $\omega_c=\frac{3}{2} \frac{c}{\rho} \gamma^3$}
  \label{fig:synchrotronspectra}
\end{figure}

From the expressions for the effective magnetic field and the energy of the particles we know the corresponding synchrotron spectra, i.e., the intensity of emitted radiation as a function of frequency $\mathrm{d} I / \mathrm{d}\omega$, see Fig. \ref{fig:synchrotronspectra}. This can be expressed as
\begin{equation}
 \label{eq:SynchrotronFunction}
 \mathrm{d}I = \frac{4}{9} \frac{ e^3 H_{\text{eff}} \omega_c}{mc^2} \left( \frac{9\sqrt{3}}{8 \pi} F\left( \omega \right)   \right)\mathrm{d}\omega ,
\end{equation}
where the integral over all $\omega$ of the expression in the parenthesis is $1$. 
To control the computational costs and the resolution of the output we can alter the number of super-photons to emit per super-particle and time step. This could be one or several super-photons per super-particle and time step or it could be less, choosing at random to emit only from a subset of super-particles. Furthermore, one can for this purpose control the number of real photons represented by a super-photon, both individually and at runtime.

To test if a particle is to emit a photon we pick a frequency $\omega_{\text{emit}}$ randomly from a certain distribution. To reduce the numerical noise we take this distribution to have characteristics similar to the synchrotron spectra, so frequencies where the synchrotron spectra is large are generated more often (to compensate for this, the probability of emission will be divided by the value of the distribution). This is achieved with the simplified function
\begin{equation}
  S(x)= \left\{
  \begin{array}{rl}
    4/3 x^{1/3} & \text{if } x < a ,\\
    7/9 e^{-x} & \text{if } x > a ,
  \end{array} \right.
\end{equation}
which has a similar asymptotic behaviour as the synchrotron spectra. The constant $a\approx 0.69021 $ is determined so that $\int_0^{\infty} S(x) = 1$. For this function  we can analytically calculate
\begin{equation}
  P(\omega) = \int_0^{\omega} S(x) \mathrm{d} x
\end{equation}
and invert it to find $\omega(P)$. We then generate $\omega_{\text{emit}}$ by $\omega_{\text{emit}}= \omega(R')$ where $R' \in [0:1]$ is a uniform (pseudo)random number. The number of photons to emit is then given by
\begin{equation}
\mathrm{d}N = \frac{\mathrm{d}I/\mathrm{d}\omega(\omega_{\text{emit}})}{\hbar \omega_{\text{emit}} S(\omega_{\text{emit}}/\omega_c)} \Delta t
\end{equation}
where the synchrotron function $\mathrm{d}I/\mathrm{d}\omega (\omega)$ is given by Eq. (\ref{eq:SynchrotronFunction}). With this often being a number less than $1$, emission is determined by comparing to a (pseudo)random number $R\in [0:1]$.

Thus, the steps employed in the computation of the synchrotron photons are
\begin{enumerate}
  \item calculate $H_{\text{eff}}$ from the change in momentum (limiting $\Delta p_{\perp}/p$ to 1),
  \item calculate the typical frequency $\omega_c$ from $H_{\text{eff}}$,
  \item generate a frequency for emission, $\omega_{\text{emit}}=\omega(R')$,
  \item calculate the distribution value $S(\omega_{\text{emit}}/\omega_c)$ of the generated frequency, 
  \item calculate probability of emission, $\mathrm{d}N$, using $\omega_{\text{emit}}$ and $S(\omega_{\text{emit}}/\omega_c)$, and 
  \item determine emission by comparing $\mathrm{d}N$ to random $R \in [0:1]$.
\end{enumerate}

\subsection{Limitations}
As mentioned, the method calculates the emission using data from only two neighbouring time steps for each particle. Thus, there is no interference between radiation from different time steps. However, we argued that this interference can be neglected for laser-plasma interaction problems in regimes where the radiation reaction force is of importance. 

The method uses the classical expression for radiation emission and is therefore not valid in the extreme case where a full QED treatment is needed, i.e. cases where the quantum parameter $\chi$ is close to or more than $1$. \cite{Lobet2013,ridgers2014modelling} It still has a very large span of validity with $\chi < 1$ for laser intensities up to $10^{24} \, \mathrm{W/cm}^2$. This means that the method can be used in simulations of forthcoming high intensity laser systems.

\section{Comparison and benchmarking}
Cyclotron motion is central for the developed method and, by construction, it samples the correct synchrotron spectra for an ultra-relativistic particle moving in an constant external magnetic field. 

\begin{figure}
\includegraphics[width=0.45\textwidth]{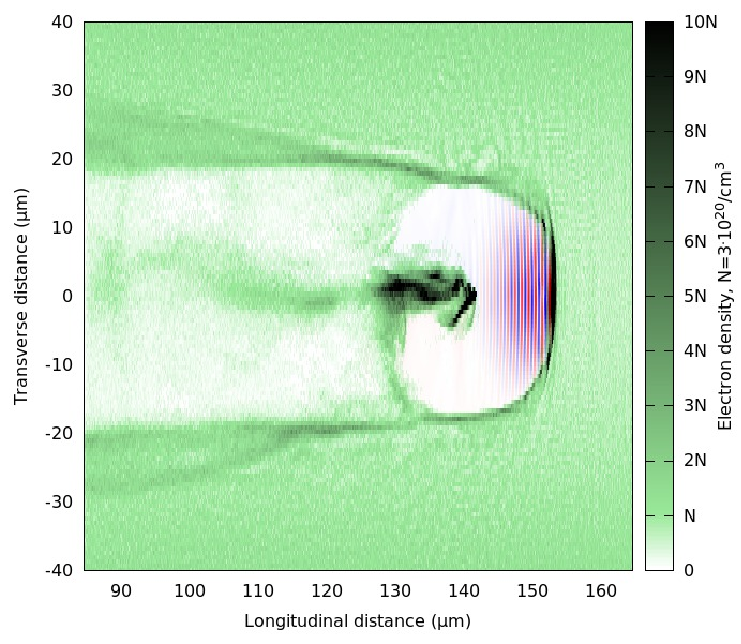}
  \caption{Wakefield simulation, green-black showing the electron density and red-blue the y-component of E-field in the laser. The picture is taken at $0.54 ps$ when the laser has moved $\approx 160 \mu m$ into the plasma}
  \label{fig:simulation}
\end{figure}

We test the method using a laser wakefield setup, seen Fig.\ \ref{fig:simulation}. A linearly polarized laser with a duration of $20\, \mathrm{fs}$, a diameter of $8 \, \mu\mathrm{m}$ (both given as FWHM), and an energy of $10^4 J$ is shot against a hydrogen plasma. The plasma has a density of $N = 3 \times 10^{20}\, \mathrm{cm}^{-3}$ along the laser optical axis, increasing parabolically by a factor $2$ at a radius of $40 \, \mu\mathrm{m}$. The simulation box is $80 \, \mu\mathrm{m} \times 80 \, \mu\mathrm{m} \times 20 \, \mu\mathrm{m}$ distributed on $512 \times 128 \times 64$ cells with $2$ particles per cell, comoving with the laser pulse. Electrons are injected into the wakefield by a plasma density profile, which increases linearly from $0$ to $3 N$ in $10 \, \mu\mathrm{m}$, then down to $N$ in another $10 \, \mu\mathrm{m}$. The laser collides with the plasma at $x=0$ with a laser intensity of $7.8 \times 10^{23} \, \mathrm{W/cm}^2$ (calculated using FWHM values). 
As expected, the electrons (green to black) are accelerated upon collision with the laser (blue and red) and driven around and through the laser pulse. Some electrons are trapped behind the pulse and are accelerated to very high energies, $\gamma \approx 10^4$. These electrons oscillate behind the laser, emitting radiation in the laser propagation direction. 

\begin{figure}
  \includegraphics[width=0.45\textwidth]{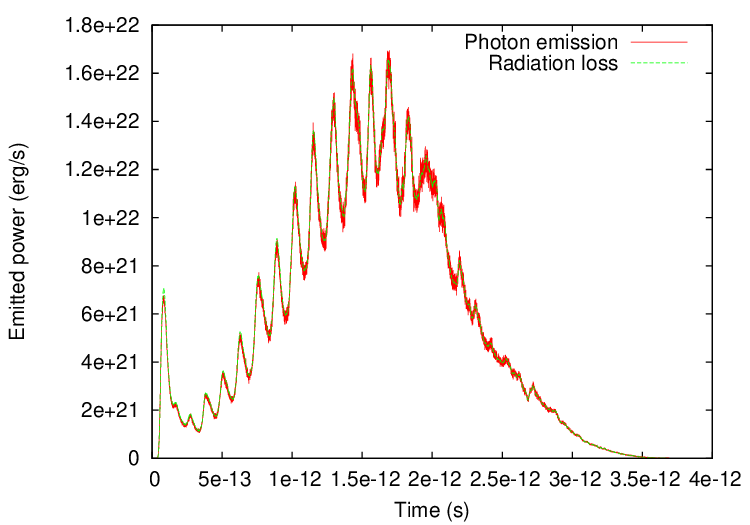}
  \caption{Emitted power due to synchrotron radiation (red/whole line) (as simulated by the described method) compared to the energy lost due to the radiation reaction (green/dashed) (as calculated using the LL formula) for a laser travelling a plasma and forming a wakefield.}
  \label{fig:RRandsynchrotronemission}
\end{figure}

We calculate the energy loss due to the radiation reaction force for the particles using the Landau-Lifshitz expression (\ref{eq:LLformOfRRForce}), and the sum of this energy loss for all particles is compared to the sum of the emitted synchrotron radiation for all time steps. As seen in Fig. \ref{fig:RRandsynchrotronemission} there is a very good agreement between these two independent calculations, throughout all stages of this simulation.

\begin{figure}
  \includegraphics[width=0.35\textwidth]{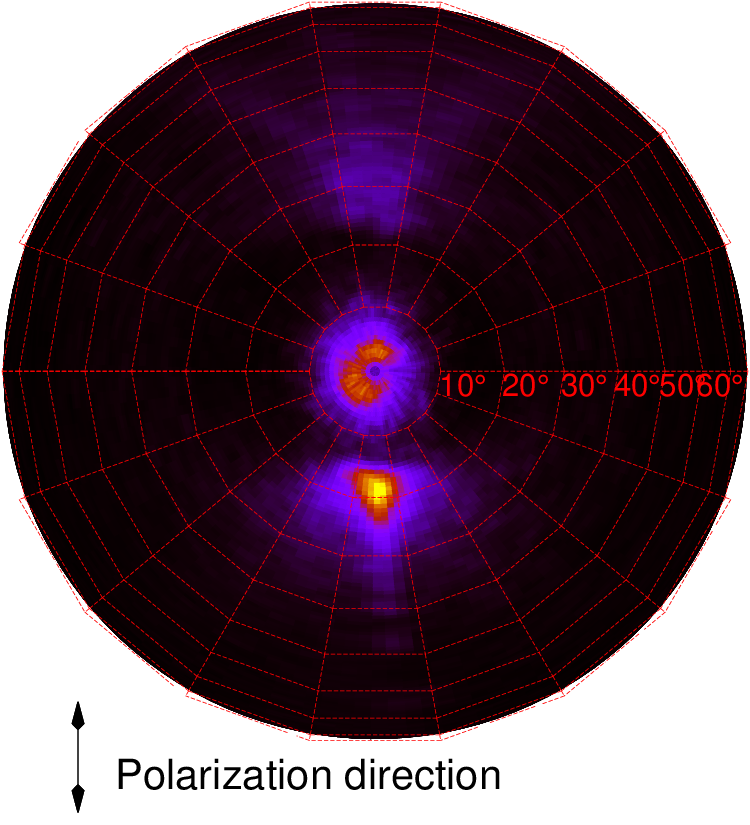}
  \caption{Typical angular distribution of emitted energy in a wakefield simulation. Plot of front hemisphere around the laser propagation direction.}
  \label{fig:angular}
\end{figure}

Looking at a typical angular distribution of the emitted energy, Fig. \ref{fig:angular}, 
we see one contribution in the center, and another part at a greater angle from the center. The central one is due to the accelerated electrons in the electron bunch, which are oscillating about the laser propagation direction. 
The radiation at a greater angle is due to the acceleration of the electrons in the laser front. There is a clear dependence of the laser polarization in the azimuthal angle of this, with the polarization of the laser being in the up-down direction.

\section{Conclusions}
We have developed a simple, yet accurate, method of emitting high energy photons in a particle-in-cell simulation during runtime. Considering a particle to be in an instantaneous circular motion we use the effective magnetic field needed to produce such a motion to sample the synchrotron spectra with a Monte-Carlo method. Furthermore, we do a rigorous comparison to the mechanism of radiation reaction, which is incorporated as a classical correction to the equations of motion through the Landau-Lifshitz model (this can easily be extended to the quantum regime \cite{QEDPIC}). The simulated emissions agree very well with the energy loss due to radiation reaction in a laser-wakefield simulation and show an expected angular distribution. All the calculations have been performed for moderate to very high intensities, the latter relevant for next generation laser experiments. It has a large validity span with applications in simulations of future high intensity laser facilities, though now valid in the extreme, full QED case. finally, we have also been able to find a method to reduce the noise in the computation of radiation losses for laser-plasma interactions. 

\section*{Acknowledgment}
This research was supported by the Swedish Research Council Grants \# 2010-3727 and 2012-3320. The simulations were performed on resources provided by the Swedish National Infrastructure for Computing (SNIC)  at High Performance Computing Center North (HPC2N).

\bibliography{References}
\bibliographystyle{unsrt}

\end{document}